\documentclass[aps,pre,reprint,superscriptaddress]{revtex4-2}

\usepackage{lineno}
\usepackage{graphicx}
\usepackage{makeidx}
\usepackage{tabularx}
\usepackage{dcolumn}

\usepackage{amsfonts}
\usepackage{amssymb}
\usepackage{amsmath}

\begin{document}

\title{Discrete equations and auto-traveling kinks of the $\phi^6$ model}
\author{H. Susanto}
\email{hadi.susanto@yandex.com}
\affiliation{Department of Mathematics, Khalifa University, PO Box 127788, Abu Dhabi, United Arab Emirates}

\author{N. Karjanto}
\email{natanael@skku.edu}
\affiliation{Department of Mathematics, University College, Sungkyunkwan University, Natural Science Campus, 2066 Seobu-ro, Jangan-gu, Suwon 16419, Gyeonggi-do, Republic of Korea}

\date{\today}

\begin{abstract}
We study the $\phi^{6}$ model and derive two broad classes of lattice discretizations that admit static, translationally invariant kinks; that is, stationary kink profiles that can be centered at an arbitrary position relative to the lattice. These discretizations are constructed using a one-dimensional map, $\phi_{n+1}=F(\phi_{n})$, which provides a direct and systematic algorithm for generating such models. Numerical computations for two representative cases show that the discrete kinks do not possess internal modes, consistent with the continuum theory, although an additional high-frequency mode may appear above the phonon band. We also show that generic discretizations of the $\phi^{6}$ model do not support static kink solutions. Instead, the resulting dynamics produce auto-traveling and self-accelerating kinks that propagate at the maximal group velocity while continuously emitting radiation.
\end{abstract}

\maketitle


\section{Introduction} \label{introduction}

Spatially discrete systems arise in many physical contexts, including dislocation dynamics in crystals and guided-wave propagation in inhomogeneous optical media~\cite{kivshar2003optical,braun2004frenkel,kevrekidis2009discrete,chong2018coherent}. Of particular interest is the behavior of traveling solitary waves in such lattices~\cite{kevrekidis2011non}, whose properties differ markedly from those of solutions of continuous homogeneous partial differential equations (PDEs). In continuum models, traveling waves can often be generated from static ones using spatial translational invariance or additional symmetries such as Galilean invariance~\cite{yang2010nonlinear}. In contrast, discrete systems do not necessarily support freely propagating localized waves. This restriction is commonly attributed to the Peierls--Nabarro barrier, an energetic obstacle that impedes movement between lattice sites~\cite{sepulchre2003energy,braun2004frenkel,al2017peierls,jenkinson2015onsite,naether2011peierls}.

Discrete systems also arise as finite-difference approximations of PDEs~\cite{ames1992numerical}. An important question in this setting is whether a numerical discretization preserves qualitative features of the underlying continuum model, in particular the ability of solitary waves to propagate without radiative energy loss, which may otherwise lead to deceleration and trapping. A well-known example is the Ablowitz--Ladik equation, a discrete analogue of the nonlinear Schr\"odinger equation that supports mobile solitary waves~\cite{ablowitz1975nonlinear,ablowitz1976nonlinear,ablowitz2007inverse}, even though it does not correspond to a physical system. More broadly, a class of discrete Schrödinger equations supporting static solitary waves with translational invariance was identified in~\cite{pelinovsky2006translationally}. Related constructions have been developed for nonlinear Klein--Gordon-type equations, including discrete $\phi^{4}$ models~\cite{barashenkov2005translationally} and the discrete sine-Gordon equation~\cite{barashenkov2008exceptional}. Further special discretizations of the $\phi^{4}$ equation with additional symmetries or conservation laws (e.g., momentum or energy) appear in~\cite{speight1994kink,kevrekidis2003class,dmitriev2005discrete,dmitriev2006exact}; see also the review~\cite{dmitriev2019discrete}.

Within the class of nonlinear Klein–Gordon-type equations, only discrete $\phi^{4}$ and sine-Gordon models with symmetric kinks have been studied in detail. Higher-order field theories such as the $\phi^{6}$ model have received comparatively little attention, despite their relevance to first-order phase transitions and related applications in materials science and condensed-matter physics~\cite{saxena2019higher}. The $\phi^{6}$ model is of particular interest because it supports asymmetric kinks without internal vibrational modes. Although such modes were once thought to be necessary for multi-bounce resonances in kink--antikink collisions, Dorey \emph{et al.}~\cite{dorey2011kink} demonstrated resonant scattering in the $\phi^{6}$ model. A collective coordinate approximation of the scattering processes was derived in \cite{gani2014kink}. Its statistical mechanics also differ substantially from those of the $\phi^{4}$ theory, leading to a quasi-exactly solvable structure~\cite{behera1980classical,bruce1980structural}. These features highlight the distinct behavior of higher-order field theories.

Discrete analogues of the $\phi^{6}$ equation were examined in~\cite{rakhmatullina2018non}, where two discretizations were derived through a discretized first integral~\cite{speight1994kink,kevrekidis2003class}. Another discretization, based on a discrete Bogomolny–Prasad–Sommerfield (BPS) construction, was proposed in~\cite{saadatmand2024phonons}. These models admit kinks that can be centered at any point between lattice sites. The present work develops additional discrete $\phi^{6}$ equations using the one-dimensional map method introduced in~\cite{barashenkov2005translationally,barashenkov2008exceptional}. We show that, aside from these exceptional cases, generic discretizations of the $\phi^{6}$ model do not support static kink solutions. Instead, they produce auto-traveling kinks that propagate and accelerate until they reach a velocity limited by the maximum group velocity of linear plane waves (the speed of sound). Beyond this threshold, accelerating kinks shed radiation in the form of shock-like wave trains. We emphasize that the absence of static kinks reported here is not a consequence of kink asymmetry or unequal vacuum energies, but rather of discretization effects that do not arise from a standard variational/Hamiltonian formulation and which destroy the stationary heteroclinic structure even when the continuum $\phi^6$ potential has degenerate minima.

The paper is organized as follows. Section~\ref{model} outlines the one-dimensional map method and derives exceptional discrete $\phi^{6}$ models. Section~\ref{numerics} presents numerical computations of static kink profiles and their stability. Section~\ref{auto} examines the mechanism and dynamics of auto-traveling kinks. Conclusions are given in Sec.~\ref{conclusion}.

\section{Model} \label{model}

The $\phi^6$ model in $(1 + 1)$-dimensional spacetime is described by the equation
\begin{equation}
\phi_{tt} = \phi_{xx} - \frac{dU}{d\phi} = \phi_{xx} - \phi(1 - \phi^2)(1 - 3\phi^2),
\label{continuumlimit}
\end{equation}
where $\phi(x,t)$ is a real-valued scalar field, subscripts denote partial differentiation, and $U(\phi) = \frac{1}{2} \phi^2 (1 - \phi^2)^2$ is the potential function. A natural discrete analogue of this model can be expressed as
\begin{equation}
\ddot{\phi}_n = \frac{\phi_{n+1} - 2\phi_n + \phi_{n-1}}{h^2} + f(\phi_{n-1}, \phi_n, \phi_{n+1}),
\label{discrete}
\end{equation}
where $h$ is the lattice spacing, $\phi_n(t) = \phi(x_n, t)$, and $x_n = hn$. In the continuum limit $h \to 0$, the function $f$ must reduce to the nonlinear term in Eq.~\eqref{continuumlimit}:
\begin{equation}
f(\phi, \phi, \phi) = -\phi(1 - \phi^2)(1 - 3\phi^2) = -\phi + 4\phi^3 - 3\phi^5.
\end{equation}
The central question is how to construct $f(\phi_{n-1}, \phi_n, \phi_{n+1})$ such that the discrete model admits a time-dependent kink solution with translational invariance.

Following~\cite{barashenkov2005translationally}, the key to achieving this lies in formulating the nonlinearity $f$ such that the stationary equation of Eq.~\eqref{discrete} can be derived from a one-dimensional map $\phi_{n+1} = F(\phi_n)$. This approach is based on the first integral of the time-independent equation of Eq.~\eqref{continuumlimit}, $\phi_{xx} = \frac{dU}{d\phi}$, which yields $\phi_x^2 = 2U(\phi)$ or $\phi_x = H(\phi, \phi)$, where
\begin{equation}
H(\phi, \phi) = \sqrt{2U(\phi)} = \phi(1 - \phi^2).
\end{equation}
The one-dimensional map $\phi_{n+1} = F(\phi_n)$ can be constructed by discretizing the first integral as
\begin{equation}
\phi_{n+1} - \phi_n = h H(\phi_{n+1}, \phi_n).
\label{map}
\end{equation}
By squaring both sides of Eq.~\eqref{map} and subtracting it from its backward counterpart (i.e., replacing $n$ with $n-1$), we obtain the following form of an exceptional discretization~\cite{barashenkov2005translationally}:
\begin{align}
\frac{\phi_{n+1} - 2\phi_n + \phi_{n-1}}{h^2} &= \frac{H^2(\phi_{n+1}, \phi_n) - H^2(\phi_n, \phi_{n-1})}{\phi_{n+1} - \phi_{n-1}} \nonumber \\
&= -f(\phi_{n-1}, \phi_n, \phi_{n+1}). \label{discreta}
\end{align}
For symmetric $H$, i.e., $H(\phi_n, \phi_{n-1}) = H(\phi_{n-1}, \phi_n)$, the discretization in Eq.~\eqref{discreta} is nonsingular, as the numerator can be factored to cancel the denominator.

The function $f$ resides in the vector space $\mathbb{P}_5$, implying that $H^2$ must belong to $\mathbb{P}_6$. While $H$ itself could be in $\mathbb{P}_3$, this is not a strict requirement. If $p_n$ denotes a polynomial function of degree $n$, then $H(\phi, \varphi) = p_3(\phi, \varphi)$ or $H(\phi, \varphi) = \sqrt{p_6(\phi, \varphi)}$. This leads to the following symmetry and continuity conditions for $n = 1, 2$:
\begin{align}
p_{3n}(\phi, \varphi) &= p_{3n}(\varphi, \phi), \quad p_{3n}(\phi, \phi) = \phi^n (1 - \phi^2)^n.
\label{sym}
\end{align}
In the following sections, we explore these two cases separately.

\subsection{Cubic Polynomial} \label{cubicp}

For $n = 1$, the most general cubic polynomial $p_3$ is given by:
\begin{align}
p_3(\phi,\varphi) &= \frac{1}{2}(\phi + \varphi) + \alpha_1(\phi - \varphi)^2 \nonumber \\
& \quad - \frac{1}{2} \left[\alpha_2 (\phi^3 + \varphi^3) + \alpha_3 \phi \varphi (\phi + \varphi) \right],
\end{align}
where $\alpha_1$, $\alpha_2$, $\alpha_3 \in \mathbb{R}$ satisfy $\alpha_2 + \alpha_3 = 1$. Certain terms, such as $(\phi - \varphi)$ and $(\phi - \varphi)^3$, are excluded to ensure the symmetry requirements in \eqref{sym}. 

Focusing solely on this polynomial does not yield models distinct from those obtained with a hexic polynomial in complete square form, i.e., $p_6 = p_3^2$, which will be discussed in Sec. \ref{hexicp} (see Eq. \eqref{comsq}). To introduce novelty, we incorporate a conservation law $\gamma I_3 = 0$ \cite{barashenkov2005translationally} into the discretization $f$, where 
\begin{align}
I_3 \equiv p_3(\phi_{n+1},\phi_n) (\phi_n - \phi_{n - 1})& \nonumber \\
 - p_3(\phi_n,\phi_{n-1})(\phi_{n+1} - \phi_n)&. \label{conlaw}
\end{align}
Adding the conservation law $\gamma I_3 = 0$ to the right-hand side of \eqref{discreta} preserves the discretization's validity while enabling the derivation of additional models.

Let $\phi_\text{av}^{(k)}$ denote the average value between two neighboring sites and their corresponding positive integer powers:
\begin{equation}
\phi_\text{av}^{(k)} = \frac{1}{2} \left(\phi_{n + 1}^{k} + \phi_{n - 1}^{k} \right), \label{phiave}
\end{equation}
where $k = 1-4$. Using \eqref{discreta} and the conservation law \eqref{conlaw}, we derive the following quintic polynomial:
\begin{widetext}
\begin{align}
f(\phi_{n + 1},\phi_{n},\phi_{n - 1}) &= - \frac{1}{2} \left(\phi_{n} + \phi_\text{av}^{(1)} \right) 
+ \alpha_1 \phi_{n} \left(\phi_{n} + 2\phi_\text{av}^{(1)} \right) 
+ \frac{1}{2} (8\alpha_1^2 + \alpha_2 + \alpha_3) \phi_{n}^3 
- 2(2\alpha_1^2 - \alpha_2) \phi_\text{av}^{(1)} \phi_\text{av}^{(2)}  \nonumber \\
& \quad - 2 (6\alpha_1^2 - \alpha_3) \phi_\text{av}^{(1)} \phi_{n}^2 
+ 2\alpha_1 (\alpha_2 - \alpha_3) \phi_\text{av}^{(1)} \phi_{n}^3 
- \alpha_1 (2 \alpha_2 - \alpha_3) \phi_{n} \left(\phi_{n}^3 + 4 \phi_\text{av}^{(1)} \phi_\text{av}^{(2)} \right)
- \frac{1}{2} \alpha_2 \alpha_3 \phi_{n}^5  \nonumber \\
& \quad - \frac{1}{2} \left[2\alpha_1 - (8\alpha_1^2 + \alpha_2 + \alpha_3)\phi_{n} - 2\alpha_1 (\alpha_2 - \alpha_3) \phi_{n}^2 + (\alpha_2^2 + \alpha_3^2) \phi_{n}^3 + \alpha_2^2 \phi_\text{av}^{(3)} \right] \left(2\phi_\text{av}^{(2)} + \phi_{n + 1} \phi_{n - 1} \right) \nonumber \\
& \quad - \frac{1}{2} \alpha_3 (2\alpha_2 + \alpha_3) \phi_\text{av}^{(1)} \phi_{n}^2 \left(\phi_{n}^2 + 2 \phi_\text{av}^{(2)} \right)
+ \frac{1}{2} \alpha_2 \left(2\alpha_1 - \alpha_3 \phi_{n} \right) \left(2\phi_\text{av}^{(4)} + 2\phi_\text{av}^{(2)} \phi_{n + 1} \phi_{n - 1} + \phi_{n + 1}^2 \phi_{n - 1}^2 \right) \nonumber \\
& \quad - \gamma \left(\phi_{n}^2 - \phi_{n + 1} \phi_{n - 1} \right) 
+ 2 \gamma \alpha_1 \left[\phi_\text{av}^{(1)}\left(3 \phi_{n}^2 + \phi_{n + 1} \phi_{n - 1} \right) - \phi_\text{av}^{(2)} \phi_{n} - \phi_{n}^3 - 2 \phi_{n + 1} \phi_{n} \phi_{n - 1} \right] \nonumber \\
& \quad - \gamma \alpha_2 \left(\phi_\text{av}^{(1)} \phi_{n}^3 + \phi_\text{av}^{(2)} \phi_{n + 1} \phi_{n - 1} - \phi_\text{av}^{(3)} \phi_{n} - \phi_{n}^4 \right)\nonumber\\
&\quad + \gamma \alpha_3 \left[\phi_\text{av}^{(1)} \phi_{n}^3 + \phi_\text{av}^{(2)} \phi_{n}^2 - \left(\phi_\text{av}^{(1)} + \phi_{n} \right) \phi_{n + 1} \phi_{n} \phi_{n - 1} \right]. \label{H3quintic}
\end{align}	
\end{widetext}
This defines the first general class of exceptional discrete $\phi^6$ models.

\subsection{Hexic Polynomial} \label{hexicp}

For $n = 2$, the most general hexic polynomial $p_6$ is constructed as:
\begin{widetext}
\begin{align}
p_6(\phi,\varphi) &= \phi \varphi - \beta_2(\phi - \varphi)^2 - 2 \left[\beta_4 (\phi^4 + \varphi^4) + \beta_5 \phi \varphi (\phi^2 + \varphi^2) + \beta_6 \phi^2 \varphi^2 \right] + \beta_7 (\phi^6 + \varphi^6) + \beta_8 \phi \varphi (\phi^4 + \varphi^4) \nonumber \\ 
& \quad  + \beta_9 \phi^2 \varphi^2 (\phi^2 + \varphi^2) + \beta_{10} \phi^3 \varphi^3  + \beta_{11} (\phi^3 + \varphi^3) + \beta_{12} \phi \varphi (\phi + \varphi) + \beta_{13} (\phi^5 + \varphi^5) + \beta_{14} \phi \varphi (\phi^3 + \varphi^3) \nonumber \\
& \quad + \beta_{15} \phi^2 \varphi^2 (\phi + \varphi), \label{genhex} 
\end{align}
\end{widetext}
where $2\beta_4 + 2\beta_5 + \beta_6 = 1$, $2\beta_7 + 2\beta_8 + 2\beta_9 + \beta_{10} = 1$, $\beta_{11} + \beta_{12} = 0$, and $\beta_{13} + \beta_{14} + \beta_{15} = 0$. A special case arises when the hexic polynomial is a complete square, i.e., $p_6 = p_3^2$, where $p_3$ is the cubic polynomial from Sec. \ref{cubicp} with $\gamma = 0$. This relationship is expressed as:
\begin{equation}
p_6 = p_3^2 = \sum_{i = 0}^{3} \sum_{j = 0}^{6 - 2i} a_{ij} \phi^{i} \varphi^{i} \left(\phi^{6 - 2i - j} + \varphi^{6 - 2i - j} \right), \label{comsq}
\end{equation}
with coefficients $a_{ij}$, $i = 0-3$, $j = 0-6$, provided in Appendix \ref{app}.

Substituting \eqref{genhex} into \eqref{discreta} yields the nonlinearity:
\begin{widetext}	
\begin{align}
f(\phi_{n + 1},\phi_{n},\phi_{n - 1}) &= -\phi_n + 2\beta_2 \left(\phi_\text{av}^{(1)} - \phi_{n} \right) + 8 \beta_4 \phi_\text{av}^{(1)} \phi_\text{av}^{(2)} 
+ 2 \beta_5 \phi_n^3 + 4 \beta_6 \phi_\text{av}^{(1)} \phi_n^2 - \beta_8 \phi_n^5 \nonumber \\
& \quad + \left(2\beta_5 \phi_{n} - 2 \beta_7 \phi_\text{av}^{(3)} - \beta_{10} \phi_n^3 - \beta_{11} - \beta_{15} \phi_{n}^2 \right) \left(2\phi_\text{av}^{(2)} + \phi_{n + 1} \phi_{n - 1} \right) \nonumber \\
& \quad  - \left(\beta_8 \phi_n + \beta_{13} \right) \left( 2 \phi_\text{av}^{(4)} + 2 \phi_\text{av}^{(2)} \phi_{n + 1} \phi_{n - 1} + \phi_{n + 1}^2 \phi_{n - 1}^2 \right) 
- 2 \beta_9 \phi_\text{av}^{(1)} \phi_n^2 \left(2 \phi_\text{av}^{(2)} + \phi_n^2 \right) \nonumber \\
& \quad - \beta_{12} \phi_{n} \left(2\phi_\text{av}^{(1)} + \phi_{n} \right) 
- \beta_{14} \phi_n \left(4 \phi_\text{av}^{(1)} \phi_\text{av}^{(2)} + \phi_{n}^3 \right) 
- 2\beta_{15} \phi_\text{av}^{(1)} \phi_{n}^3. \label{H6quintic}
\end{align}	
\end{widetext}
This defines the second general class of exceptional discrete $\phi^6$ models. Note that model \eqref{H3quintic} cannot be derived from \eqref{H6quintic} unless $\gamma = 0$ and the coefficients satisfy the relations provided in Appendix \ref{app}.

The following section will illustrate the general models \eqref{H3quintic} and \eqref{H6quintic} for specific parameter values.

\section{Numerical Illustrations} \label{numerics}

Rakhmatullina et al. \cite{rakhmatullina2018non} proposed two special discrete models that conserve physical quantities. Their momentum-conserving model (Eq. (18) of \cite{rakhmatullina2018non}) can be derived from the discrete model \eqref{H6quintic} by setting $\beta_j = 0$ for $j \neq 6, 10$ and $\beta_6 = \beta_{10} = 1$, yielding:
\begin{equation}
f = -\phi_{n} + 4\phi_n^2 \phi_\text{av}^{(1)} - \phi_{n}^3 \left(2\phi_\text{av}^{(2)} + \phi_{n + 1} \phi_{n - 1}\right).
\label{ex1}
\end{equation}
However, their energy-conserving model cannot be directly retrieved. The closest approximation is obtained by setting $\beta_5 = 1/3$, $\beta_6 = 3/8$, $\beta_8 = 2/9$ or $1/6$, $\beta_9 = 5/36$, $\beta_{10} = 4/9$, $\beta_{11} = -1/6$, $\beta_{12} = 1/2$, and $\beta_j = 0$ otherwise in \eqref{H6quintic}. Nonetheless, the resulting equation lacks the following terms:
\begin{equation}
\frac{-16}{144}\phi_\text{av}^{(5)}, \quad \frac{8}{144}\phi_\text{av}^{(3)}(9 - 15\phi_n^2),
\label{miss}
\end{equation}
which appear in Eq. (15) of \cite{rakhmatullina2018non}. These terms were omitted in \eqref{genhex} because they violate the symmetry requirement \eqref{sym}.

We will study model \eqref{ex2} and also consider the discrete model \eqref{H3quintic} with $\alpha_1 = \alpha_2 = 0$ and $\alpha_3 = 1$, which gives:
\begin{equation}
\begin{split}
f = &\phi_\text{av}^{(1)}\left(-\frac{1}{2} + \phi_{n}^2 \left(2 - \frac{1}{2}\phi_{n}^2 - \phi_\text{av}^{(2)}\right)\right) \\
&+ \frac{1}{2}(\phi_{n} - \phi_{n}^3)\left(-1 + 2\phi_\text{av}^{(2)} + \phi_{n + 1} \phi_{n - 1}\right) \\
&- \gamma \left(\phi_{n}^2 - \phi_{n + 1} \phi_{n - 1} - \phi_\text{av}^{(1)} \phi_{n}^3 - \phi_\text{av}^{(2)} \phi_{n}^2 \right. \\
&+ \left. (\phi_\text{av}^{(1)} + \phi_{n}) \phi_{n + 1} \phi_{n} \phi_{n - 1} \right).
\end{split}
\label{ex2}
\end{equation}
We solve Eqs. \eqref{ex1}--\eqref{ex2} for kink solutions. The continuous $\phi^6$ model \eqref{continuumlimit} admits the kink:
\begin{equation}
\Phi_K(x) = \sqrt{\frac{1 + \tanh x}{2}}.
\label{kink}
\end{equation}
We numerically continue this kink for the discrete models. After obtaining a solution, $\Phi_{K,n}$, we compute its linear spectrum by substituting $\phi_n = \Phi_{K,n} + v_n e^{\lambda t}$ into \eqref{discrete} and linearizing about small $v$. This yields the eigenvalue problem:
\begin{equation}
\lambda^2 \overline{v} = J \overline{v},
\end{equation}
where $J$ is the Jacobian matrix about the discrete kink $\Phi_{K,n}$. In the continuum limit, $J = \partial_{xx} - U_{\phi\phi}|_{\phi = \Phi_K}$, with a discrete spectrum $\lambda^2 = 0$ and a continuous spectrum $\lambda^2 < -1$ \cite{dorey2011kink}.

\begin{figure}[tbhp]
\centering
\includegraphics[width=0.5\textwidth]{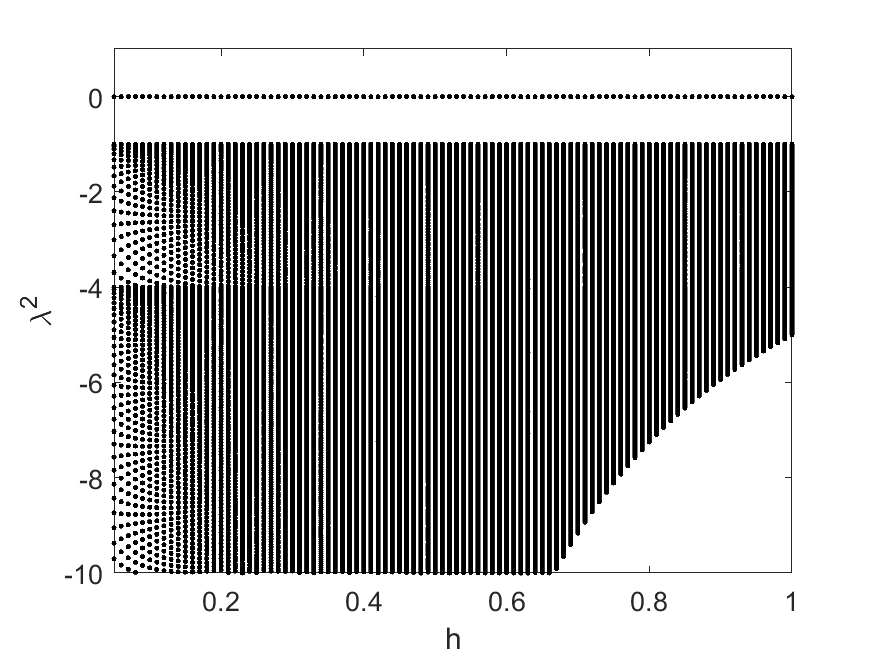} \\
\includegraphics[width=0.5\textwidth]{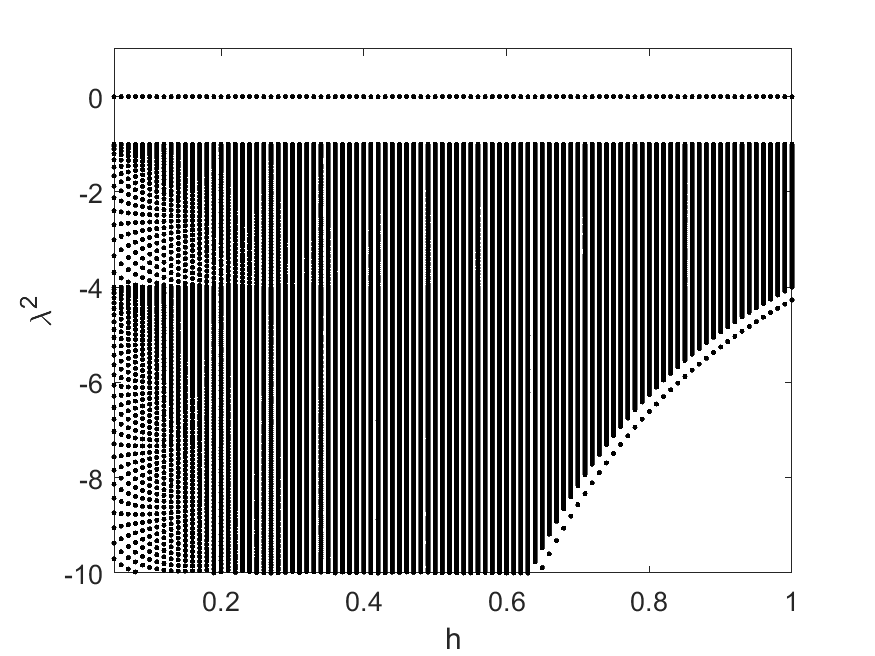}
\caption{Linear spectrum of a kink solution from the discrete models \eqref{ex1} (top panel) and \eqref{ex2} (bottom panel) as a function of the discretization $h$.}
\label{fig:spec}
\end{figure}

Figure \ref{fig:spec} shows the linear spectrum of the kink solutions for models \eqref{ex1} and \eqref{ex2}. The presence of an eigenvalue $\lambda = 0$ confirms the translational invariance of the kink. Notably, neither model exhibits internal modes between the zero eigenvalue and the edge of the continuous spectrum at $\lambda^2 = -1$, consistent with the continuum limit. However, model \eqref{ex2} exhibits a high-frequency internal mode below the lower edge of the continuous spectrum.

Similar high-frequency modes have been reported for kinks in discrete $\phi^4$ models with momentum conservation \cite{dmitriev2005discrete}. These modes correspond to out-of-phase oscillations of the kink's shape \cite{braun1997kink} and do not radiate because their harmonics are not resonant with the continuous spectrum. As a result, they can store significant energy. For further analysis, see \cite{derks2007stability}.

\section{Auto-Traveling Kinks} \label{auto}

In general, non-standard discretizations of \eqref{continuumlimit} destroy the existence of static kinks. Consider, for example:
\begin{equation}
f = -\phi_n + 4\left[\phi_\text{av}^{(1)}\right]^3 - 3 \phi_n^5.
\label{g1}
\end{equation}
In the continuum limit $h \to 0$, the discrete equation \eqref{discrete} with \eqref{g1} becomes:
\begin{equation}
\phi_{tt} = \phi_{xx} - \phi + 4\phi^3 - 3\phi^5 + \frac{h^2}{12} \left(\partial_x^4 \phi + 72\phi^2 \phi_{xx}\right).
\label{cont1}
\end{equation}
We demonstrate that, for sufficiently small $h$, Eq.~\eqref{cont1} does not admit a static kink solution that is a smooth perturbation of the continuum kink $\Phi_K$ in Eq.~\eqref{kink}; specifically, no solution exists with
\begin{equation}
\phi(x) = \Phi_K(x) + h^2 u(x) + \dots.
\label{exp}
\end{equation}
and bounded derivatives up to fourth order, having the same asymptotic states as $x\to\pm\infty$.

Substituting this expansion into \eqref{cont1} yields the inhomogeneous linear equation:
\begin{equation}
-\mathcal{L}u = \frac{1}{12} \partial_x^4 \Phi_K + 6\Phi_K^2 \partial_x^2 \Phi_K,
\label{corr}
\end{equation}
where $\mathcal{L} = \partial_{xx} - 1 + 12\Phi_K^2 - 15\Phi_K^4$ is a self-adjoint operator. For \eqref{corr} to have a localized solution, its right-hand side must be orthogonal to the null space of $\mathcal{L}$, which is spanned by $\partial_x \Phi_K$. However, we find:
\begin{equation}
\int_{-\infty}^{\infty} \Phi_K^2 \partial_x^2 \Phi_K \, \partial_x \Phi_K \, dx = -\frac{1}{24} \neq 0.
\label{corr2}
\end{equation}
Thus, no bounded solution $u$ exists for Eq.~\eqref{corr}, indicating that the perturbation breaks the regular intersection between the stable and unstable manifolds of $\phi=0$ and $\phi=1$ for any static solution that is a smooth deformation of $\Phi_{K}$. In other words, a discrete heteroclinic orbit that is $C^{4}$-close to the continuum kink cannot persist for small $h$. Nevertheless, note that, because Eq.~\eqref{cont1} contains a small parameter multiplying the highest derivative, this argument applies to smooth perturbations of $\Phi_{K}$ and does not exclude the possibility of more singular kink-like solutions with internal layers or rapidly varying structure on the lattice scale. However, our numerical simulations did not reveal any such static configurations; instead, the dynamics produce self-accelerating kinks, as described below.

\begin{figure}[tbhp]
\centering
\includegraphics[width=0.5\textwidth]{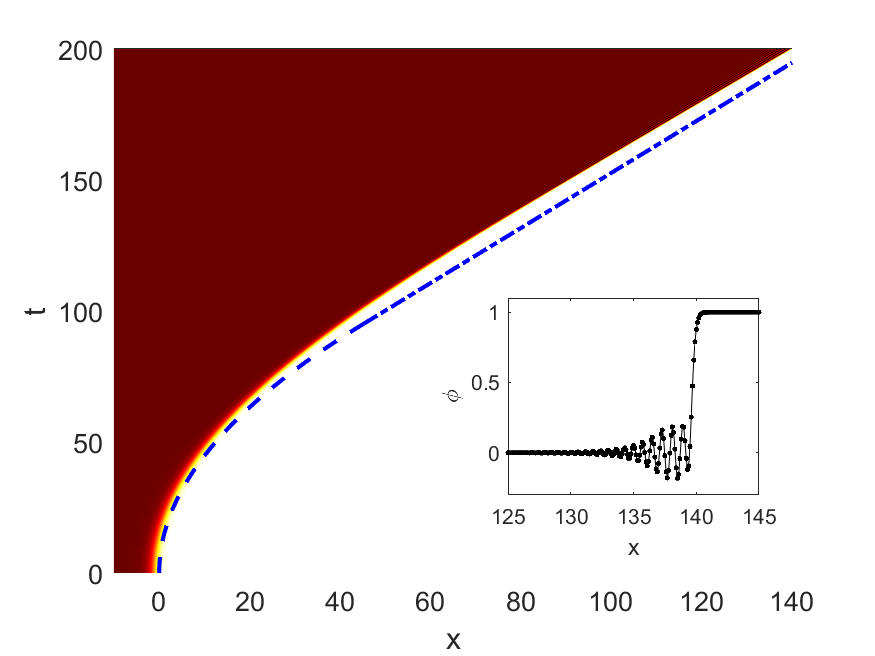}
\caption{Top view of $\phi_n$ for $h = 0.1$ in the $(x = nh, t)$-plane. The dashed curve represents \eqref{post}, while the dash-dotted line corresponds to the critical speed \eqref{maxv}. The inset shows $\phi_n$ at $t = 200$.}
\label{fig:accel}
\end{figure}

The absence of a static kink suggests the existence of a time-dependent solution. We consider a traveling kink ansatz $\Phi_K(x - x_0(t))$, where $x_0(t)$ is the kink's position. Using a variational approach \cite{dawes2013variational}, we project the resulting equation onto the kink's Goldstone mode and integrate over $x$, yielding:
\begin{equation}
x_0 = \frac{h^2}{2} t^2,
\label{post}
\end{equation}
assuming the kink is initially at rest. This implies that the kink self-accelerates. Similar results, where perturbations accelerate asymmetric kinks, were reported before in \cite{weigel2017vacuum,romanczukiewicz2017could}. However, the acceleration is bounded by a critical speed derived from the linear dispersion relation:
\begin{equation}
\left. \frac{\partial \omega}{\partial k} \right|_{k = k_c} = \sqrt{1 + \frac{h^2}{2} - \frac{h^3/2 + 2h}{\sqrt{4 + h^2}}}.
\label{maxv}
\end{equation}
For $h = 0.1$, this critical speed is approximately $0.95$. Figure \ref{fig:accel} shows the kink's dynamics, confirming the self-acceleration and the eventual saturation at the critical speed.

The kink continuously radiates energy as it approaches the critical speed, as seen in the inset of Fig.\ \ref{fig:accel}. This behavior contrasts with non-Hamiltonian discrete $\phi^4$ models, where self-accelerating kinks exist despite the presence of static kinks \cite{dmitriev2005discrete,dmitriev2006standard}. In our case, the absence of static kinks drives the self-acceleration, akin to driven Frenkel-Kontorova models \cite{braun2004frenkel}, see Appendix \ref{fk}.


It is also useful to note that, up to the higher-derivative correction, the right-hand side of Eq.~(\ref{cont1}) may be formally written as
\[
(1+6h^{2}\phi^{2})\,[\phi_{xx}-U'(\phi)],
\]
where $U$ may be interpreted as an effective potential. In this representation, the homogeneous states $\phi=0,\pm1$ remain exact equilibria, but the projection onto the translational mode acquires a nonvanishing contribution of order $h^{2}$, consistent with Eq.~(\ref{corr2}). This provides an intuitive interpretation of the acceleration mechanism, whereby the kink experiences a small effective dynamical bias. We emphasize that this viewpoint is heuristic: our conclusions ultimately rest on the breakdown of the stationary heteroclinic structure rather than on any variational formulation. This mechanism is distinct from false-vacuum acceleration, where a kink is driven by a true energy difference $V(\text{false})-V(\text{true})$ (see, e.g., \cite{kiselev1998forced}); that scenario does not apply here because the homogeneous equilibria remain degenerate and no conserved energy functional ranks them.

\section{Conclusion} \label{conclusion}

We have derived a broad family of discrete higher-order $\phi^{6}$ Klein--Gordon models that admit static, translationally invariant kinks. Numerical computations verified the associated zero eigenvalue of the linearization, confirming translational invariance, and showed the absence of internal modes within the finite band gap, consistent with the continuum $\phi^{6}$ theory. We also found that generic, non-exceptional discretizations do not support static kink solutions. Instead, they generate self-accelerating kinks that approach a maximal propagation speed set by the sonic limit, beyond which further acceleration produces sustained radiative emission.

Several avenues remain open for future work. A natural direction is to determine sliding velocities for exceptional discretizations and characterize how they depend on system parameters. It is also important to quantify the radiation amplitude and spectral content associated with self-accelerating kinks, and to understand how these radiative processes influence long-time dynamics. Another open problem concerns the interaction of kinks in both exceptional and non-exceptional models, including kink–antikink collisions and the possibility of resonance phenomena. Finally, extensions of the map-based construction to other higher-order field theories, multi-component lattices, and appropriate continuum limits may provide further insight into how discrete acceleration mechanisms and solitary-wave structures arise from the underlying lattice framework.

\section*{Acknowledgement}
HS acknowledges support by Khalifa University through  
	a Competitive Internal Research Awards Grant (No.\ 8474000413/CIRA-2021-065) and Research \& Innovation Grants (No.\ 8474000617/RIG-S-2023-031 and No.\ 8474000789/RIG-S-2024-070). The authors also thank the anonymous referees for their constructive comments and feedback that improved the manuscript. 

\appendix
\section{Coefficients in Eq.\ \eqref{comsq}}\label{app}

The coefficients of the complete square relation between \( p_6 \) and \( p_3 \) in Eq.\ \eqref{comsq} are as follows:
\begin{widetext}
	\begin{align*}
		a_{00} &= \beta_7 = \frac{1}{4} \alpha_2^2, \,
		a_{01} = \beta_{13} = -\alpha_1 \alpha_2, \,
		a_{02} = -2 \beta_4 = \frac{1}{2}(2\alpha_1^2 - \alpha_2), \\
		a_{03} &= \beta_{11} = \alpha_1, \,
		a_{04} = -\beta_2 = \frac{1}{4}, \,
		a_{05} = 0 = a_{06}, \,
		a_{10} = \beta_8 = \frac{1}{2} \alpha_2 \alpha_3, \\
		a_{11} &= \beta_{14} = \alpha_1 (2\alpha_2 - \alpha_3) = \alpha_1 (3\alpha_2 - 1), \,
		a_{12} = - 2 \beta_5 = - \frac{1}{2} (8\alpha_1^2 + 1), \,
		a_{13} = \beta_{12} = - \alpha_1, \\
		a_{14} &= \frac{1}{2}(1 + 2\beta_2) = \frac{1}{4}, \,
		a_{20} = \beta_9 = \frac{1}{4} \alpha_3 (2\alpha_2 + \alpha_3) = \frac{1}{4} (1 - \alpha_2^2), \\
		a_{21} &= \beta_{15} = \alpha_1 (\alpha_3 - \alpha_2), \,
		a_{22} = - \beta_6 = \frac{1}{2} (6\alpha_1^2 - \alpha_3), \,
		a_{30} = \frac{1}{2} \beta_{10} = \frac{1}{4} (\alpha_2^2 + \alpha_3^2).
	\end{align*}
\end{widetext}

\section{Driven Frenkel-Kontorova Model}\label{fk}

The driven Frenkel-Kontorova model is described by the equation:
\begin{align}
	\ddot{\phi}_n = \frac{\phi_{n+1} - 2\phi_n + \phi_{n-1}}{h^2} - \sin\phi_{n} + \gamma. \label{fke}
\end{align}
In the continuum limit \( h \to 0 \) and \( \gamma = 0 \), the model admits a kink solution:
\begin{equation}
	\phi_{FK}(x,t) = 4\tan^{-1}\exp x. \label{kin}
\end{equation}
This static kink persists for \( 0 < h \ll 1 \) and \( \gamma < \gamma_c \), where the critical drive \( \gamma_c \) is derived using exponential asymptotics as \cite{king2001asymptotics, carpio2003depinning}:
\begin{equation}
	\gamma_c \sim \frac{356.1}{h^2}\exp\left[-\frac{\pi^2}{2\sinh^{-1}(h/2)} \right].
\end{equation}
For \( h \gg 1 \), the kink solution \eqref{kin} also exists for \( \gamma < \gamma_c \), with the leading-order approximation of the critical drive given by \cite{derks2007stability}:
\begin{align}
	\gamma_c = 1 - \frac{2\pi}{h^2} + \mathcal{O}(1/h^4).
\end{align}
When \( \gamma > \gamma_c \), the heteroclinic manifolds connecting the fixed points \( \phi_n = \arcsin\gamma \) and \( \phi_n = 2\pi + \arcsin\gamma \) vanish, implying the absence of static kinks.

The dynamics of the kink have been extensively studied in \cite{king2001asymptotics, carpio2003depinning} for the parabolic version of the Frenkel-Kontorova model \eqref{fke}, where the left-hand side of the equation is \( \dot{\phi}_n \) instead of \( \ddot{\phi}_n \). In this case, the kink travels with a constant average velocity.

For the hyperbolic case \eqref{fke}, when \( \gamma > \gamma_c \), the kink accelerates over time, similar to the auto-travelling kinks discussed in Sec.\ \ref{auto}. Furthermore, by seeking a travelling kink solution of the form \( \phi_{FK}(x - x_0, t) \), we find that the kink's center \( x_0(t) \) in the limit \( h \to 0 \) is given by:
\begin{align}
	x_0 = -\frac{\pi}{8}\gamma t^2.
\end{align}


Note that in the driven Frenkel--Kontorova model above, the acceleration of kinks results from a drive-induced bias that eliminates static kink solutions. In contrast, in the discrete $\phi^6$ models considered in this work, static kinks can be destroyed even in the absence of vacuum energy bias, as a consequence of discretization effects that break the stationary heteroclinic structure and do not rely on any assumption of an underlying variational or Hamiltonian structure.

\bibliography{references}

\providecommand{\noopsort}[1]{}\providecommand{\singleletter}[1]{#1}%
\begin{thebibliography}{38}%
\makeatletter
\providecommand \@ifxundefined [1]{%
 \@ifx{#1\undefined}
}%
\providecommand \@ifnum [1]{%
 \ifnum #1\expandafter \@firstoftwo
 \else \expandafter \@secondoftwo
 \fi
}%
\providecommand \@ifx [1]{%
 \ifx #1\expandafter \@firstoftwo
 \else \expandafter \@secondoftwo
 \fi
}%
\providecommand \natexlab [1]{#1}%
\providecommand \enquote  [1]{``#1''}%
\providecommand \bibnamefont  [1]{#1}%
\providecommand \bibfnamefont [1]{#1}%
\providecommand \citenamefont [1]{#1}%
\providecommand \href@noop [0]{\@secondoftwo}%
\providecommand \href [0]{\begingroup \@sanitize@url \@href}%
\providecommand \@href[1]{\@@startlink{#1}\@@href}%
\providecommand \@@href[1]{\endgroup#1\@@endlink}%
\providecommand \@sanitize@url [0]{\catcode `\\12\catcode `\$12\catcode
  `\&12\catcode `\#12\catcode `\^12\catcode `\_12\catcode `\%12\relax}%
\providecommand \@@startlink[1]{}%
\providecommand \@@endlink[0]{}%
\providecommand \url  [0]{\begingroup\@sanitize@url \@url }%
\providecommand \@url [1]{\endgroup\@href {#1}{\urlprefix }}%
\providecommand \urlprefix  [0]{URL }%
\providecommand \Eprint [0]{\href }%
\providecommand \doibase [0]{https://doi.org/}%
\providecommand \selectlanguage [0]{\@gobble}%
\providecommand \bibinfo  [0]{\@secondoftwo}%
\providecommand \bibfield  [0]{\@secondoftwo}%
\providecommand \translation [1]{[#1]}%
\providecommand \BibitemOpen [0]{}%
\providecommand \bibitemStop [0]{}%
\providecommand \bibitemNoStop [0]{.\EOS\space}%
\providecommand \EOS [0]{\spacefactor3000\relax}%
\providecommand \BibitemShut  [1]{\csname bibitem#1\endcsname}%
\let\auto@bib@innerbib\@empty
\bibitem [{\citenamefont {Kivshar}\ and\ \citenamefont
  {Agrawal}(2003)}]{kivshar2003optical}%
  \BibitemOpen
  \bibfield  {author} {\bibinfo {author} {\bibfnamefont {Y.~S.}\ \bibnamefont
  {Kivshar}}\ and\ \bibinfo {author} {\bibfnamefont {G.~P.}\ \bibnamefont
  {Agrawal}},\ }\href@noop {} {\emph {\bibinfo {title} {Optical solitons: from
  fibers to photonic crystals}}}\ (\bibinfo  {publisher} {Academic Press},\
  \bibinfo {year} {2003})\BibitemShut {NoStop}%
\bibitem [{\citenamefont {Braun}\ and\ \citenamefont
  {Kivshar}(2004)}]{braun2004frenkel}%
  \BibitemOpen
  \bibfield  {author} {\bibinfo {author} {\bibfnamefont {O.~M.}\ \bibnamefont
  {Braun}}\ and\ \bibinfo {author} {\bibfnamefont {Y.~S.}\ \bibnamefont
  {Kivshar}},\ }\href@noop {} {\emph {\bibinfo {title} {The Frenkel-Kontorova
  model: concepts, methods, and applications}}}\ (\bibinfo  {publisher}
  {Springer, Berlin},\ \bibinfo {year} {2004})\BibitemShut {NoStop}%
\bibitem [{\citenamefont {Kevrekidis}(2009)}]{kevrekidis2009discrete}%
  \BibitemOpen
  \bibfield  {author} {\bibinfo {author} {\bibfnamefont {P.~G.}\ \bibnamefont
  {Kevrekidis}},\ }\href@noop {} {\emph {\bibinfo {title} {The discrete
  nonlinear Schr{\"o}dinger equation: mathematical analysis, numerical
  computations and physical perspectives}}},\ Vol.\ \bibinfo {volume} {232}\
  (\bibinfo  {publisher} {Springer Science \& Business Media},\ \bibinfo {year}
  {2009})\BibitemShut {NoStop}%
\bibitem [{\citenamefont {Chong}\ and\ \citenamefont
  {Kevrekidis}(2018)}]{chong2018coherent}%
  \BibitemOpen
  \bibfield  {author} {\bibinfo {author} {\bibfnamefont {C.}~\bibnamefont
  {Chong}}\ and\ \bibinfo {author} {\bibfnamefont {P.~G.}\ \bibnamefont
  {Kevrekidis}},\ }\href@noop {} {\emph {\bibinfo {title} {Coherent structures
  in granular crystals: From experiment and modelling to computation and
  mathematical analysis}}}\ (\bibinfo  {publisher} {Springer},\ \bibinfo {year}
  {2018})\BibitemShut {NoStop}%
\bibitem [{\citenamefont {Kevrekidis}(2011)}]{kevrekidis2011non}%
  \BibitemOpen
  \bibfield  {author} {\bibinfo {author} {\bibfnamefont {P.~G.}\ \bibnamefont
  {Kevrekidis}},\ }\href@noop {} {\bibfield  {journal} {\bibinfo  {journal}
  {IMA Journal of Applied Mathematics}\ }\textbf {\bibinfo {volume} {76}},\
  \bibinfo {pages} {389} (\bibinfo {year} {2011})}\BibitemShut {NoStop}%
\bibitem [{\citenamefont {Yang}(2010)}]{yang2010nonlinear}%
  \BibitemOpen
  \bibfield  {author} {\bibinfo {author} {\bibfnamefont {J.}~\bibnamefont
  {Yang}},\ }\href@noop {} {\emph {\bibinfo {title} {Nonlinear waves in
  integrable and nonintegrable systems}}}\ (\bibinfo  {publisher} {SIAM},\
  \bibinfo {year} {2010})\BibitemShut {NoStop}%
\bibitem [{\citenamefont {Sepulchre}(2003)}]{sepulchre2003energy}%
  \BibitemOpen
  \bibfield  {author} {\bibinfo {author} {\bibfnamefont {J.~A.}\ \bibnamefont
  {Sepulchre}},\ }in\ \href@noop {} {\emph {\bibinfo {booktitle} {Localization
  and Energy Transfer in Nonlinear Systems}}}\ (\bibinfo  {publisher} {World
  Scientific},\ \bibinfo {year} {2003})\ pp.\ \bibinfo {pages}
  {102--129}\BibitemShut {NoStop}%
\bibitem [{\citenamefont {Al~Khawaja}\ \emph {et~al.}(2017)\citenamefont
  {Al~Khawaja}, \citenamefont {Al-Marzoug},\ and\ \citenamefont
  {Bahlouli}}]{al2017peierls}%
  \BibitemOpen
  \bibfield  {author} {\bibinfo {author} {\bibfnamefont {U.}~\bibnamefont
  {Al~Khawaja}}, \bibinfo {author} {\bibfnamefont {S.}~\bibnamefont
  {Al-Marzoug}},\ and\ \bibinfo {author} {\bibfnamefont {H.}~\bibnamefont
  {Bahlouli}},\ }\href@noop {} {\bibfield  {journal} {\bibinfo  {journal}
  {Communications in Nonlinear Science and Numerical Simulation}\ }\textbf
  {\bibinfo {volume} {46}},\ \bibinfo {pages} {74} (\bibinfo {year}
  {2017})}\BibitemShut {NoStop}%
\bibitem [{\citenamefont {Jenkinson}\ and\ \citenamefont
  {Weinstein}(2015)}]{jenkinson2015onsite}%
  \BibitemOpen
  \bibfield  {author} {\bibinfo {author} {\bibfnamefont {M.}~\bibnamefont
  {Jenkinson}}\ and\ \bibinfo {author} {\bibfnamefont {M.~I.}\ \bibnamefont
  {Weinstein}},\ }\href@noop {} {\bibfield  {journal} {\bibinfo  {journal}
  {Nonlinearity}\ }\textbf {\bibinfo {volume} {29}},\ \bibinfo {pages} {27}
  (\bibinfo {year} {2015})}\BibitemShut {NoStop}%
\bibitem [{\citenamefont {Naether}\ \emph {et~al.}(2011)\citenamefont
  {Naether}, \citenamefont {Vicencio},\ and\ \citenamefont
  {Johansson}}]{naether2011peierls}%
  \BibitemOpen
  \bibfield  {author} {\bibinfo {author} {\bibfnamefont {U.}~\bibnamefont
  {Naether}}, \bibinfo {author} {\bibfnamefont {R.~A.}\ \bibnamefont
  {Vicencio}},\ and\ \bibinfo {author} {\bibfnamefont {M.}~\bibnamefont
  {Johansson}},\ }\href@noop {} {\bibfield  {journal} {\bibinfo  {journal}
  {Physical Review E}\ }\textbf {\bibinfo {volume} {83}},\ \bibinfo {pages}
  {036601} (\bibinfo {year} {2011})}\BibitemShut {NoStop}%
\bibitem [{\citenamefont {Ames}(1992)}]{ames1992numerical}%
  \BibitemOpen
  \bibfield  {author} {\bibinfo {author} {\bibfnamefont {W.~F.}\ \bibnamefont
  {Ames}},\ }\href@noop {} {\emph {\bibinfo {title} {Numerical Methods for
  Partial Differential Equations}}}\ (\bibinfo  {publisher} {Academic, New
  York},\ \bibinfo {year} {1992})\BibitemShut {NoStop}%
\bibitem [{\citenamefont {Ablowitz}\ and\ \citenamefont
  {Ladik}(1975)}]{ablowitz1975nonlinear}%
  \BibitemOpen
  \bibfield  {author} {\bibinfo {author} {\bibfnamefont {M.~J.}\ \bibnamefont
  {Ablowitz}}\ and\ \bibinfo {author} {\bibfnamefont {J.~F.}\ \bibnamefont
  {Ladik}},\ }\href@noop {} {\bibfield  {journal} {\bibinfo  {journal} {Journal
  of Mathematical Physics}\ }\textbf {\bibinfo {volume} {16}},\ \bibinfo
  {pages} {598} (\bibinfo {year} {1975})}\BibitemShut {NoStop}%
\bibitem [{\citenamefont {Ablowitz}\ and\ \citenamefont
  {Ladik}(1976)}]{ablowitz1976nonlinear}%
  \BibitemOpen
  \bibfield  {author} {\bibinfo {author} {\bibfnamefont {M.}~\bibnamefont
  {Ablowitz}}\ and\ \bibinfo {author} {\bibfnamefont {J.}~\bibnamefont
  {Ladik}},\ }\href@noop {} {\bibfield  {journal} {\bibinfo  {journal} {Journal
  of Mathematical Physics}\ }\textbf {\bibinfo {volume} {17}},\ \bibinfo
  {pages} {1011} (\bibinfo {year} {1976})}\BibitemShut {NoStop}%
\bibitem [{\citenamefont {Ablowitz}\ \emph {et~al.}(2007)\citenamefont
  {Ablowitz}, \citenamefont {Biondini},\ and\ \citenamefont
  {Prinari}}]{ablowitz2007inverse}%
  \BibitemOpen
  \bibfield  {author} {\bibinfo {author} {\bibfnamefont {M.~J.}\ \bibnamefont
  {Ablowitz}}, \bibinfo {author} {\bibfnamefont {G.}~\bibnamefont {Biondini}},\
  and\ \bibinfo {author} {\bibfnamefont {B.}~\bibnamefont {Prinari}},\
  }\href@noop {} {\bibfield  {journal} {\bibinfo  {journal} {Inverse Problems}\
  }\textbf {\bibinfo {volume} {23}},\ \bibinfo {pages} {1711} (\bibinfo {year}
  {2007})}\BibitemShut {NoStop}%
\bibitem [{\citenamefont {Pelinovsky}(2006)}]{pelinovsky2006translationally}%
  \BibitemOpen
  \bibfield  {author} {\bibinfo {author} {\bibfnamefont {D.~E.}\ \bibnamefont
  {Pelinovsky}},\ }\href@noop {} {\bibfield  {journal} {\bibinfo  {journal}
  {Nonlinearity}\ }\textbf {\bibinfo {volume} {19}},\ \bibinfo {pages} {2695}
  (\bibinfo {year} {2006})}\BibitemShut {NoStop}%
\bibitem [{\citenamefont {Barashenkov}\ \emph {et~al.}(2005)\citenamefont
  {Barashenkov}, \citenamefont {Oxtoby},\ and\ \citenamefont
  {Pelinovsky}}]{barashenkov2005translationally}%
  \BibitemOpen
  \bibfield  {author} {\bibinfo {author} {\bibfnamefont {I.~V.}\ \bibnamefont
  {Barashenkov}}, \bibinfo {author} {\bibfnamefont {O.~F.}\ \bibnamefont
  {Oxtoby}},\ and\ \bibinfo {author} {\bibfnamefont {D.~E.}\ \bibnamefont
  {Pelinovsky}},\ }\href@noop {} {\bibfield  {journal} {\bibinfo  {journal}
  {Physical Review E}\ }\textbf {\bibinfo {volume} {72}},\ \bibinfo {pages}
  {035602(R)} (\bibinfo {year} {2005})}\BibitemShut {NoStop}%
\bibitem [{\citenamefont {Barashenkov}\ and\ \citenamefont {van
  Heerden}(2008)}]{barashenkov2008exceptional}%
  \BibitemOpen
  \bibfield  {author} {\bibinfo {author} {\bibfnamefont {I.~V.}\ \bibnamefont
  {Barashenkov}}\ and\ \bibinfo {author} {\bibfnamefont {T.~C.}\ \bibnamefont
  {van Heerden}},\ }\href@noop {} {\bibfield  {journal} {\bibinfo  {journal}
  {Physical Review E}\ }\textbf {\bibinfo {volume} {77}},\ \bibinfo {pages}
  {036601} (\bibinfo {year} {2008})}\BibitemShut {NoStop}%
\bibitem [{\citenamefont {Speight}\ and\ \citenamefont
  {Ward}(1994)}]{speight1994kink}%
  \BibitemOpen
  \bibfield  {author} {\bibinfo {author} {\bibfnamefont {J.~M.}\ \bibnamefont
  {Speight}}\ and\ \bibinfo {author} {\bibfnamefont {R.~S.}\ \bibnamefont
  {Ward}},\ }\href@noop {} {\bibfield  {journal} {\bibinfo  {journal}
  {Nonlinearity}\ }\textbf {\bibinfo {volume} {7}},\ \bibinfo {pages} {475}
  (\bibinfo {year} {1994})}\BibitemShut {NoStop}%
\bibitem [{\citenamefont {Kevrekidis}(2003)}]{kevrekidis2003class}%
  \BibitemOpen
  \bibfield  {author} {\bibinfo {author} {\bibfnamefont {P.~G.}\ \bibnamefont
  {Kevrekidis}},\ }\href@noop {} {\bibfield  {journal} {\bibinfo  {journal}
  {Physica D: Nonlinear Phenomena}\ }\textbf {\bibinfo {volume} {183}},\
  \bibinfo {pages} {68} (\bibinfo {year} {2003})}\BibitemShut {NoStop}%
\bibitem [{\citenamefont {Dmitriev}\ \emph {et~al.}(2005)\citenamefont
  {Dmitriev}, \citenamefont {Kevrekidis},\ and\ \citenamefont
  {Yoshikawa}}]{dmitriev2005discrete}%
  \BibitemOpen
  \bibfield  {author} {\bibinfo {author} {\bibfnamefont {S.}~\bibnamefont
  {Dmitriev}}, \bibinfo {author} {\bibfnamefont {P.}~\bibnamefont
  {Kevrekidis}},\ and\ \bibinfo {author} {\bibfnamefont {N.}~\bibnamefont
  {Yoshikawa}},\ }\href@noop {} {\bibfield  {journal} {\bibinfo  {journal}
  {Journal of Physics A: Mathematical and General}\ }\textbf {\bibinfo {volume}
  {38}},\ \bibinfo {pages} {7617} (\bibinfo {year} {2005})}\BibitemShut
  {NoStop}%
\bibitem [{\citenamefont {Dmitriev}\ \emph
  {et~al.}(2006{\natexlab{a}})\citenamefont {Dmitriev}, \citenamefont
  {Kevrekidis}, \citenamefont {Yoshikawa},\ and\ \citenamefont
  {Frantzeskakis}}]{dmitriev2006exact}%
  \BibitemOpen
  \bibfield  {author} {\bibinfo {author} {\bibfnamefont {S.~V.}\ \bibnamefont
  {Dmitriev}}, \bibinfo {author} {\bibfnamefont {P.~G.}\ \bibnamefont
  {Kevrekidis}}, \bibinfo {author} {\bibfnamefont {N.}~\bibnamefont
  {Yoshikawa}},\ and\ \bibinfo {author} {\bibfnamefont {D.~J.}\ \bibnamefont
  {Frantzeskakis}},\ }\href@noop {} {\bibfield  {journal} {\bibinfo  {journal}
  {Physical Review E}\ }\textbf {\bibinfo {volume} {74}},\ \bibinfo {pages}
  {046609} (\bibinfo {year} {2006}{\natexlab{a}})}\BibitemShut {NoStop}%
\bibitem [{\citenamefont {Dmitriev}\ and\ \citenamefont
  {Kevrekidis}(2019)}]{dmitriev2019discrete}%
  \BibitemOpen
  \bibfield  {author} {\bibinfo {author} {\bibfnamefont {S.~V.}\ \bibnamefont
  {Dmitriev}}\ and\ \bibinfo {author} {\bibfnamefont {P.~G.}\ \bibnamefont
  {Kevrekidis}},\ }in\ \href@noop {} {\emph {\bibinfo {booktitle} {A Dynamical
  Perspective on the $\phi^4$ Model}}}\ (\bibinfo  {publisher} {Springer},\
  \bibinfo {year} {2019})\ pp.\ \bibinfo {pages} {111--136}\BibitemShut
  {NoStop}%
\bibitem [{\citenamefont {Saxena}\ \emph {et~al.}(2019)\citenamefont {Saxena},
  \citenamefont {Christov},\ and\ \citenamefont {Khare}}]{saxena2019higher}%
  \BibitemOpen
  \bibfield  {author} {\bibinfo {author} {\bibfnamefont {A.}~\bibnamefont
  {Saxena}}, \bibinfo {author} {\bibfnamefont {I.~C.}\ \bibnamefont
  {Christov}},\ and\ \bibinfo {author} {\bibfnamefont {A.}~\bibnamefont
  {Khare}},\ }in\ \href@noop {} {\emph {\bibinfo {booktitle} {A Dynamical
  Perspective on the $\phi^4$ Model}}}\ (\bibinfo  {publisher} {Springer},\
  \bibinfo {year} {2019})\ pp.\ \bibinfo {pages} {253--279}\BibitemShut
  {NoStop}%
\bibitem [{\citenamefont {Dorey}\ \emph {et~al.}(2011)\citenamefont {Dorey},
  \citenamefont {Mersh}, \citenamefont {Romanczukiewicz},\ and\ \citenamefont
  {Shnir}}]{dorey2011kink}%
  \BibitemOpen
  \bibfield  {author} {\bibinfo {author} {\bibfnamefont {P.}~\bibnamefont
  {Dorey}}, \bibinfo {author} {\bibfnamefont {K.}~\bibnamefont {Mersh}},
  \bibinfo {author} {\bibfnamefont {T.}~\bibnamefont {Romanczukiewicz}},\ and\
  \bibinfo {author} {\bibfnamefont {Y.}~\bibnamefont {Shnir}},\ }\href@noop {}
  {\bibfield  {journal} {\bibinfo  {journal} {Physical Review Letters}\
  }\textbf {\bibinfo {volume} {107}},\ \bibinfo {pages} {091602} (\bibinfo
  {year} {2011})}\BibitemShut {NoStop}%
\bibitem [{\citenamefont {Gani}\ \emph {et~al.}(2014)\citenamefont {Gani},
  \citenamefont {Kudryavtsev},\ and\ \citenamefont {Lizunova}}]{gani2014kink}%
  \BibitemOpen
  \bibfield  {author} {\bibinfo {author} {\bibfnamefont {V.~A.}\ \bibnamefont
  {Gani}}, \bibinfo {author} {\bibfnamefont {A.~E.}\ \bibnamefont
  {Kudryavtsev}},\ and\ \bibinfo {author} {\bibfnamefont {M.~A.}\ \bibnamefont
  {Lizunova}},\ }\href@noop {} {\bibfield  {journal} {\bibinfo  {journal}
  {Physical Review D}\ }\textbf {\bibinfo {volume} {89}},\ \bibinfo {pages}
  {125009} (\bibinfo {year} {2014})}\BibitemShut {NoStop}%
\bibitem [{\citenamefont {Behera}\ and\ \citenamefont
  {Khare}(1980)}]{behera1980classical}%
  \BibitemOpen
  \bibfield  {author} {\bibinfo {author} {\bibfnamefont {S.~N.}\ \bibnamefont
  {Behera}}\ and\ \bibinfo {author} {\bibfnamefont {A.}~\bibnamefont {Khare}},\
  }\href@noop {} {\bibfield  {journal} {\bibinfo  {journal} {Pramana}\ }\textbf
  {\bibinfo {volume} {15}},\ \bibinfo {pages} {245} (\bibinfo {year}
  {1980})}\BibitemShut {NoStop}%
\bibitem [{\citenamefont {Bruce}(1980)}]{bruce1980structural}%
  \BibitemOpen
  \bibfield  {author} {\bibinfo {author} {\bibfnamefont {A.~D.}\ \bibnamefont
  {Bruce}},\ }\href@noop {} {\bibfield  {journal} {\bibinfo  {journal}
  {Advances in Physics}\ }\textbf {\bibinfo {volume} {29}},\ \bibinfo {pages}
  {111} (\bibinfo {year} {1980})}\BibitemShut {NoStop}%
\bibitem [{\citenamefont {Rakhmatullina}\ \emph {et~al.}(2018)\citenamefont
  {Rakhmatullina}, \citenamefont {Kevrekidis},\ and\ \citenamefont
  {Dmitriev}}]{rakhmatullina2018non}%
  \BibitemOpen
  \bibfield  {author} {\bibinfo {author} {\bibfnamefont {Z.~G.}\ \bibnamefont
  {Rakhmatullina}}, \bibinfo {author} {\bibfnamefont {P.~G.}\ \bibnamefont
  {Kevrekidis}},\ and\ \bibinfo {author} {\bibfnamefont {S.~V.}\ \bibnamefont
  {Dmitriev}},\ }in\ \href@noop {} {\emph {\bibinfo {booktitle} {IOP Conference
  Series: Materials Science and Engineering}}},\ Vol.\ \bibinfo {volume} {447}\
  (\bibinfo {organization} {IOP Publishing},\ \bibinfo {year} {2018})\ p.\
  \bibinfo {pages} {012057}\BibitemShut {NoStop}%
\bibitem [{\citenamefont {Saadatmand}\ \emph {et~al.}(2024)\citenamefont
  {Saadatmand}, \citenamefont {Marjaneh}, \citenamefont {Askari},\ and\
  \citenamefont {Weigel}}]{saadatmand2024phonons}%
  \BibitemOpen
  \bibfield  {author} {\bibinfo {author} {\bibfnamefont {D.}~\bibnamefont
  {Saadatmand}}, \bibinfo {author} {\bibfnamefont {A.~M.}\ \bibnamefont
  {Marjaneh}}, \bibinfo {author} {\bibfnamefont {A.}~\bibnamefont {Askari}},\
  and\ \bibinfo {author} {\bibfnamefont {H.}~\bibnamefont {Weigel}},\
  }\href@noop {} {\bibfield  {journal} {\bibinfo  {journal} {Chaos, Solitons \&
  Fractals}\ }\textbf {\bibinfo {volume} {180}},\ \bibinfo {pages} {114550}
  (\bibinfo {year} {2024})}\BibitemShut {NoStop}%
\bibitem [{\citenamefont {Braun}\ \emph {et~al.}(1997)\citenamefont {Braun},
  \citenamefont {Kivshar},\ and\ \citenamefont {Peyrard}}]{braun1997kink}%
  \BibitemOpen
  \bibfield  {author} {\bibinfo {author} {\bibfnamefont {O.~M.}\ \bibnamefont
  {Braun}}, \bibinfo {author} {\bibfnamefont {Y.~S.}\ \bibnamefont {Kivshar}},\
  and\ \bibinfo {author} {\bibfnamefont {M.}~\bibnamefont {Peyrard}},\
  }\href@noop {} {\bibfield  {journal} {\bibinfo  {journal} {Physical Review
  E}\ }\textbf {\bibinfo {volume} {56}},\ \bibinfo {pages} {6050} (\bibinfo
  {year} {1997})}\BibitemShut {NoStop}%
\bibitem [{\citenamefont {Derks}\ \emph {et~al.}(2007)\citenamefont {Derks},
  \citenamefont {Doelman}, \citenamefont {Van~Gils},\ and\ \citenamefont
  {Susanto}}]{derks2007stability}%
  \BibitemOpen
  \bibfield  {author} {\bibinfo {author} {\bibfnamefont {G.}~\bibnamefont
  {Derks}}, \bibinfo {author} {\bibfnamefont {A.}~\bibnamefont {Doelman}},
  \bibinfo {author} {\bibfnamefont {S.}~\bibnamefont {Van~Gils}},\ and\
  \bibinfo {author} {\bibfnamefont {H.}~\bibnamefont {Susanto}},\ }\href@noop
  {} {\bibfield  {journal} {\bibinfo  {journal} {SIAM journal on applied
  dynamical systems}\ }\textbf {\bibinfo {volume} {6}},\ \bibinfo {pages} {99}
  (\bibinfo {year} {2007})}\BibitemShut {NoStop}%
\bibitem [{\citenamefont {Dawes}\ and\ \citenamefont
  {Susanto}(2013)}]{dawes2013variational}%
  \BibitemOpen
  \bibfield  {author} {\bibinfo {author} {\bibfnamefont {J.~H.~P.}\
  \bibnamefont {Dawes}}\ and\ \bibinfo {author} {\bibfnamefont
  {H.}~\bibnamefont {Susanto}},\ }\href@noop {} {\bibfield  {journal} {\bibinfo
   {journal} {Physical Review E}\ }\textbf {\bibinfo {volume} {87}},\ \bibinfo
  {pages} {063202} (\bibinfo {year} {2013})}\BibitemShut {NoStop}%
\bibitem [{\citenamefont {Weigel}(2017)}]{weigel2017vacuum}%
  \BibitemOpen
  \bibfield  {author} {\bibinfo {author} {\bibfnamefont {H.}~\bibnamefont
  {Weigel}},\ }\href@noop {} {\bibfield  {journal} {\bibinfo  {journal}
  {Physics Letters B}\ }\textbf {\bibinfo {volume} {766}},\ \bibinfo {pages}
  {65} (\bibinfo {year} {2017})}\BibitemShut {NoStop}%
\bibitem [{\citenamefont
  {Roma{\'n}czukiewicz}(2017)}]{romanczukiewicz2017could}%
  \BibitemOpen
  \bibfield  {author} {\bibinfo {author} {\bibfnamefont {T.}~\bibnamefont
  {Roma{\'n}czukiewicz}},\ }\href@noop {} {\bibfield  {journal} {\bibinfo
  {journal} {Physics Letters B}\ }\textbf {\bibinfo {volume} {773}},\ \bibinfo
  {pages} {295} (\bibinfo {year} {2017})}\BibitemShut {NoStop}%
\bibitem [{\citenamefont {Dmitriev}\ \emph
  {et~al.}(2006{\natexlab{b}})\citenamefont {Dmitriev}, \citenamefont
  {Kevrekidis},\ and\ \citenamefont {Yoshikawa}}]{dmitriev2006standard}%
  \BibitemOpen
  \bibfield  {author} {\bibinfo {author} {\bibfnamefont {S.}~\bibnamefont
  {Dmitriev}}, \bibinfo {author} {\bibfnamefont {P.}~\bibnamefont
  {Kevrekidis}},\ and\ \bibinfo {author} {\bibfnamefont {N.}~\bibnamefont
  {Yoshikawa}},\ }\href@noop {} {\bibfield  {journal} {\bibinfo  {journal}
  {Journal of Physics A: Mathematical and General}\ }\textbf {\bibinfo {volume}
  {39}},\ \bibinfo {pages} {7217} (\bibinfo {year}
  {2006}{\natexlab{b}})}\BibitemShut {NoStop}%
\bibitem [{\citenamefont {Kiselev}\ and\ \citenamefont
  {Shnir}(1998)}]{kiselev1998forced}%
  \BibitemOpen
  \bibfield  {author} {\bibinfo {author} {\bibfnamefont {V.}~\bibnamefont
  {Kiselev}}\ and\ \bibinfo {author} {\bibfnamefont {Y.~M.}\ \bibnamefont
  {Shnir}},\ }\href@noop {} {\bibfield  {journal} {\bibinfo  {journal}
  {Physical Review D}\ }\textbf {\bibinfo {volume} {57}},\ \bibinfo {pages}
  {5174} (\bibinfo {year} {1998})}\BibitemShut {NoStop}%
\bibitem [{\citenamefont {King}\ and\ \citenamefont
  {Chapman}(2001)}]{king2001asymptotics}%
  \BibitemOpen
  \bibfield  {author} {\bibinfo {author} {\bibfnamefont {J.}~\bibnamefont
  {King}}\ and\ \bibinfo {author} {\bibfnamefont {S.}~\bibnamefont {Chapman}},\
  }\href@noop {} {\bibfield  {journal} {\bibinfo  {journal} {European Journal
  of Applied Mathematics}\ }\textbf {\bibinfo {volume} {12}},\ \bibinfo {pages}
  {433} (\bibinfo {year} {2001})}\BibitemShut {NoStop}%
\bibitem [{\citenamefont {Carpio}\ and\ \citenamefont
  {Bonilla}(2003)}]{carpio2003depinning}%
  \BibitemOpen
  \bibfield  {author} {\bibinfo {author} {\bibfnamefont {A.}~\bibnamefont
  {Carpio}}\ and\ \bibinfo {author} {\bibfnamefont {L.~L.}\ \bibnamefont
  {Bonilla}},\ }\href@noop {} {\bibfield  {journal} {\bibinfo  {journal} {SIAM
  Journal on Applied Mathematics}\ }\textbf {\bibinfo {volume} {63}},\ \bibinfo
  {pages} {1056} (\bibinfo {year} {2003})}\BibitemShut {NoStop}%
\end{thebibliography}%

\end{document}